\documentclass[structabstract]{aa}  
\usepackage{graphicx}
\usepackage{txfonts}
\begin{document}
   \title{Spectrophotometric Redshifts}

   \subtitle{A New Approach to the Reduction of Noisy Spectra\\ 
and its Application to GRB090423}

   \author{M. Stefanon\inst{1}
          \and A. Fernandez-Soto\inst{2}
          \and D. Fugazza\inst{3}
          }

   \institute{Observatori Astronomic de la Universitat de Valencia, 
C/ Catedratico Agustin Escardino Benlloch 7, Paterna, E-46980-Valencia, Spain\\
   	\email{mauro.stefanon@uv.es}
         \and Instituto de Fisica de Cantabria (CSIC-UC), 
Edificio Juan Jorda, Av. los Castros s/n, E-39005-Santander, Spain\\
	\email{fsoto@ifca.unican.es}
	\and INAF - Osservatorio Astronomico di Brera, 
Via Bianchi 46, I-23807-Merate(LC), Italy\\
	\email{dino.fugazza@brera.inaf.it}
          }

\date{Submitted 13May 2010; received ; accepted }

 
\abstract
  {The measurement of redshifts for objects at the verge of
    feasibility is difficult and prone to errors. This is specially
    true in the case of almost featureless spectra, as is the case of
    GRB afterglows. They can be detected to the largest distances, and
    usually spectroscopy poses a serious problem because of their
    quick fading.}
  {We have developed a new method, close in philosophy to the
    photometric redshift technique, which can be applied to spectral
    data of very low signal-to-noise ratio. Using it we intend to
    measure redshifts while minimising the dangers posed by the usual
    extraction techniques.}
  {GRB afterglows have generally very simple optical spectra, which
    can be well described by a pure power law, over which the separate
    effects of absorption and reddening in the GRB host, the
    intergalactic medium, and our own Galaxy are superimposed. We
    model all these effects over a series of template afterglow
    spectra to produce a set of clean spectra that reproduce what
    would reach our telescope. We also model carefully the effects of
    the telescope-spectrograph combination and the properties of noise
    in the data, which are then applied on the template spectra. The
    final templates are compared to the two-dimensional spectral data,
    and the basic parameters (redshift, spectral index, Hydrogen
    absorption column), are estimated using statistical tools.}
  {We show how our method works by applying it to our data of the NIR
    afterglow of GRB090423. At $z \approx 8.2$, this was the most
    distant object ever observed. Our team took a spectrum using the
    Telescopio Nazionale Galileo, that we use in this article to
    derive its redshift and its intrinsic neutral
    Hydrogen column density.
    Our best fit yields $z=8.4^{+0.05}_{-0.03}$ and $\rm{N(HI)}<5\times10^{20}\rm{cm}^{-2}$), but with a highly non-Gaussian uncertainty including the redshift range $z \in [6.7, 8.5]$ at the 2-sigma confidence level.}
  {Our method will be useful to maximise the recovered information
    from low-quality spectra, particularly when the set of possible
    spectra is limited or easily parameterisable (as is the case in
    GRB afterglows) while at the same time ensuring an adequate
    confidence analysis.}


   \maketitle

%

\section{Introduction}

The measurement of redshifts in astronomy is one of the most important
techniques. However, it is also one that depends critically on the
quality of the available data, and because of its nature it is
sometimes difficult to attest the quality of the results, as no quantitative
error estimates are obtained. In the
optical range, the standard approach is to reduce all the available
data to a one-dimensional array, which is flux- and
wavelength-calibrated with the help of auxiliary data. In most cases
detection of emission and/or absorption lines is necessary for a valid
measurement, although in some instances only the continuum and some
basic spectral features (e.g. breaks) are needed. Even the latter is
sometimes difficult because of the paucity of photons. Ideally, in
cases of low signal-to-noise data, one would bin the spectrum in the
wavelength direction, but even this is sometimes useless. Moreover,
information is often lost in the process of extracting the spectrum.

A different approach from the informational point of view would be to
choose a model that represents the best possible fit to the available
two-dimensional spectral data. This is actually the approach used by
photometric redshift techniques, when a series of spectral energy
distributions are considered at different redshifts, and converted
into photometric data that can be compared with the available
photometry. It is also the method that has become standard in high
energy (X and gamma-ray) spectroscopy, where the models are convolved
with the instrumental response and compared to the data, instead of
the data being extracted and calibrated. In order for this kind of
approach to work, at least three conditions need to be fulfilled:

\begin{figure*}
\centering
\includegraphics[angle=0,width=16cm]{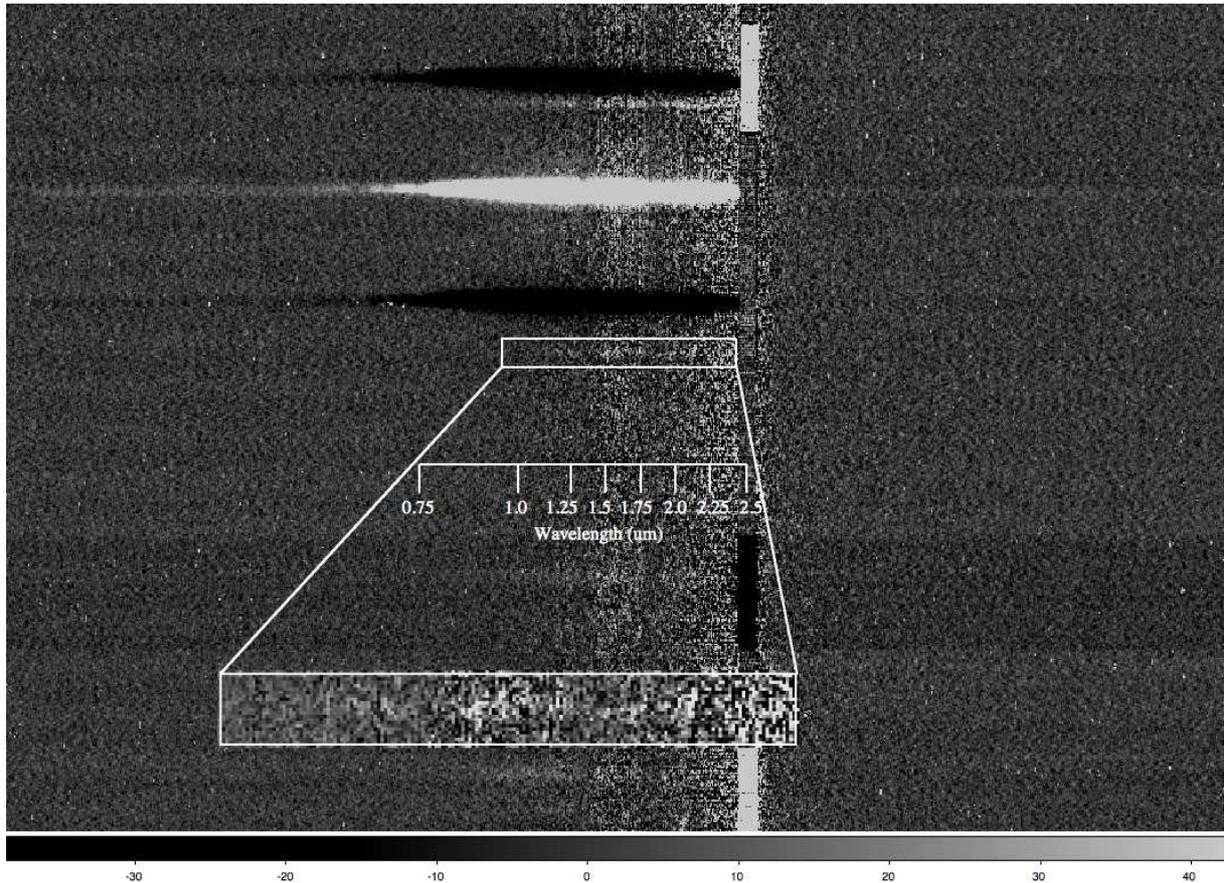}
\caption{The combined two-dimensional spectrum used as input data. The
  white rectangle identifies the region of the GRB afterglow spectrum
  that was used for the analysis. The box height is equal to $4\times
  fwhm$ corresponding to 21 pixels, ensuring that we completely
  include both wings of the gaussian. The inset shows a magnification
  of the same region.}
      \label{FigFits}
\end{figure*}

\begin{itemize}
\item The real spectrum must be included in the family of models under
  analysis. This would be relatively difficult, for instance, in the
  case of quasars or galaxies at moderate resolution, as the intrinsic
  scatter amongst different types or models is very large. However, GRB
  afterglows make for an excellent example, as their intrinsic optical
  spectrum can usually be very well approximated with a single
  power-law, where the effects of the GRB host gas and dust, the
  intergalactic medium, and our local extinction, can be superimposed.

\item The technical characteristics of the instrument must be well
  known, in order to model their effect into the simulated data. This
  includes the total wavelength-dependent efficiency, any possible
  geometric distortions in the spectral direction, and of course the
  exact position of the target in the slit image.

\item The characteristics of the noise in the CCD must also be
  modelled accurately, so that the statistical analysis will render an
  accurate estimate of the model parameters and also of the
  uncertainties associated to them.
\end{itemize}

Even though we have not mentioned it explicitly, it is of course
necessary to ensure that an accurate first reduction of the data is
performed, going from the raw individual images to the combined
two-dimensional spectrum, which constitutes the input data for our
analysis.

\section{Description of the Data}

In order to describe our work in detail, we concentrate in the
particular case for which we developed our original idea, thus we
start by describing those data.

\subsection{GRB090423 Afterglow Data}

GRB090423 was a gamma-ray burst detected by the {\it Swift} satellite
on April 23, 2009 (Krimm et al 2009). Early observations in the
optical and near infrared distinctly pointed towards the possibility
of it being a high-redshift object, when it went undetected for all
observers using visible bands, but showed as a relatively bright
near-infrared source (Tanvir et al 2009a, Cucchiara et al
2009a). Photometric data alone indicated a very high-redshift nature,
with basically zero dust absorption (Cucchiara et al 2009b, Olivares
et al 2009). Our group used the Italian 3.6m Telescopio Nazionale
Galileo, in the island of La Palma, to obtain a low-resolution
spectrum using the Amici prism with the spectrograph NICS (Oliva
2003), and measured its redshift to be $z=8.1^{+0.1}_{-0.3}$ (Thoene
et al 2009, Fernandez-Soto et al 2009, Salvaterra et al 2009). A
compatible result ($z=8.23^{+0.06}_{-0.07}$) was reached independently
by Tanvir et al (2009b, 2009c) using two sets of higher-quality data
obtained with the VLT in Chile.

The Amici spectrum covers in a single exposure the wavelength range
$0.8-2.5 \mu\rm{m}$ with very low resolution ($R \approx 50$) but very
high efficiency, and thus became the ideal choice for this kind of
analysis. We obtained a total of 128 minutes of on-target exposure
time. The exposures were dithered following the usual NIR technique,
and combined into a single two-dimensional frame, which is showed in
Figure 1. The slit was positioned with the help of a nearby star,
whose extracted spectrum will be one of the keys in our analysis.

The position of the star along the slit (measured at the reference
position X=600 in the CCD frame) is Y=765. The angular distance
between the reference star and the afterglow was $\approx 30$
arcseconds, which corresponds to 120 pixels along the slit. In order
to avoid possible issues caused by misalignments or the effect of
distortions in the focal plane, we use this distance only as a
reference, and perform a careful recentering, as described in the next
Section.

\section{Description of the Method}

Our aim will be to reproduce as perfectly as possible the spectrum of
the afterglow of GRB090423, and to reconstruct the spectral equivalent
of the wavelength-dependent \emph{point spread function} as generated
when the light passes through the atmosphere, telescope and instrument
optics, and reaches the detector. We present in this Section the
different steps that we perform to reach this objective.

\subsection{Model spectra}
We create a library of model spectra, where the basic input parameters
are three: the redshift $z$, the slope $\alpha$ in the power-law
spectral model $f_\nu \propto \lambda ^{\alpha}$, and the total
neutral Hydrogen column density in the host interstellar medium N(HI),
that produces a strong Damped Lyman Alpha (DLA) profile at the host
redshift. It is important to include this profile in the analysis,
because a dense ($\rm{N(HI)} \gtrsim 10^{21} \rm{cm}^{-2}$) DLA
profile would displace the position of the break, thus mimicking a
higher redshift. A fourth parameter, the apparent magnitude
normalization in the observed $K$ band (at the epoch of our
observations) $m_K$, will be left as a non-interesting parameter and
directly fitted to the data during the process.

\begin{figure}
\centering
\includegraphics[angle=-0,width=8cm]{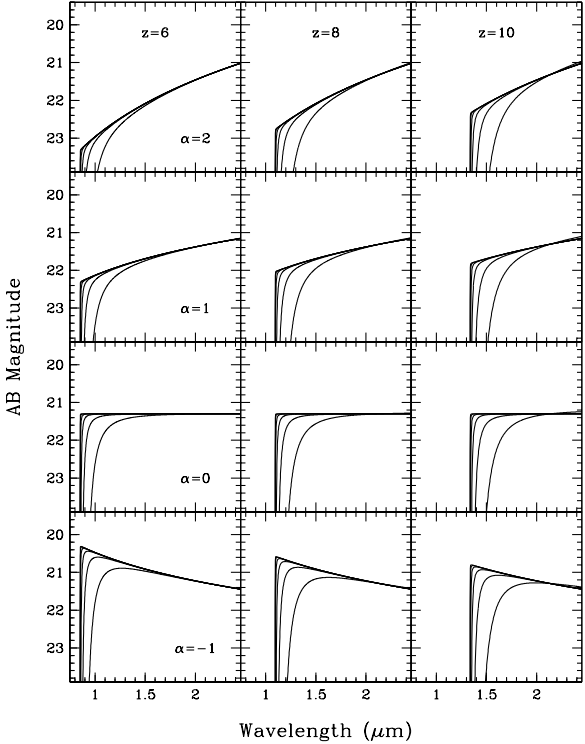}
   \caption{Some of the templates used for the analysis. Each column
     corresponds to a different redshift as labelled in the top
     panels, and each line to a different spectral slope, as labelled
     on the leftmost column. Within each panel, the different curves
     correspond to values of $\log$[N(HI)] varying from 19 to 24 in
     steps of one.
           }
      \label{FigTemps}
\end{figure}

The effect of the Intergalactic Medium (IGM) at the redshifts of interest
($z \gtrsim 6$) is very simple to include. At such a high redshift the HI
absorption is complete---within our observational capabilities---below
the Lyman-$\alpha$ line, and as such we include it in the models
(Yoshii \& Peterson 1994). The putative effects of a significatively different neutral fraction in the IGM were deliberately neglected for two reasons: on one side, the literature on GRB090423 points to a normal environment (i.e. neutral, see Tanvir et al. 2009c); on the other side some authors have shown that those effects are difficult to model and challenging to observe even with better quality data (e.g. Patel et al. 2010).

In other cases one would of course need different parameters. It could
be necessary to add dust extinction either at the host or by the Milky
Way (or both), with the amount of extinction and even its grain type
left as free parameters. We do not consider it here, because the
available afterglow photometry indicates a blue object, with an almost
complete lack of intrinsic extinction (Fernandez-Soto et al
2009 and Tanvir et al 2009c, in particular their Figure 2 where it is shown how the spectral slope is well represented by a pure power-law). Moreover, given the narrow rest-frame wavelength range that we are observing ($\lambda \approx 1200-2200 \AA$), the resolution and signal-to-noise ratio of our data, there is an almost perfect degeneracy between the amount of dust extinction and a change in spectral slope.\footnote{We used the Schlegel et al (1998) maps to include in
  the templates the effect of Milky Way dust at the level of
  E($B-V$)=0.03.}

Figure 2 shows a selection of spectral templates, sampling part of the
parameter space. The full range covered by our templates is $\alpha
\in [-1,3]$, $z \in [5,10]$, and N(HI)$\in
[10^{20},10^{24}]\rm{cm}^{-2}$. All three ranges safely include the
expected values of each variable.  All models are normalised to have
$AB_K=21.3$, a value measured by GROND (Tanvir et al 2009c) at almost exactly the
same time our observations were performed. It must be remarked,
however, that because of the possibility of slit losses affecting the
detected flux, we leave the normalisation factor for the flux as a
free parameter, as explained in the next subsection.

\begin{table}
  \caption[]{Stellar data. SDSS and 2MASS photometry were taken from
    their respective catalogues, and is expressed in their usual
    reference--AB magnitudes for SDSS, Vega-based for 2MASS.}
         \label{TabStar}
  $$
         \begin{array}{p{0.5\linewidth}l}
            \hline
            \noalign{\smallskip}
            Parameter      &  Value \\
            \noalign{\smallskip}
            \hline
            \noalign{\smallskip}
            SDSS name         & J095535.28+180903.8  \\
            R.A. (J2000)      & $09:55:35.286$  \\
            Dec (J2000)       & $+18:09:03.88$  \\
            Spectral Type     &  M4V \\
            $u_{\rm{SDSS}}$  &  22.17 \pm 0.21 \\
            $g_{\rm{SDSS}}$  &  19.27 \pm 0.01 \\
            $r_{\rm{SDSS}}$  &  17.78 \pm 0.01 \\
            $i_{\rm{SDSS}}$  &  16.54 \pm 0.01 \\
            $z_{\rm{SDSS}}$  &  15.91 \pm 0.01 \\
            $J_{\rm{2MASS}}$  &  14.52 \pm 0.03 \\
            $H_{\rm{2MASS}}$  &  13.97 \pm 0.04 \\
            $K_{\rm{2MASS}}$  &  13.77 \pm 0.05 \\
            \noalign{\smallskip}
            \hline
         \end{array}
  $$
\end{table}

\subsection{CCD and Instrumental Characteristics}

Once satisfied with the set of spectral templates, we need to
characterise the observations in terms of spectral resolution,
efficiency of the instrument at different wavelengths, noise
characterics of the detector, and position of the spectrum along both
the spectral and spatial directions.

\subsubsection{Instrumental Characteristics}

We have used an archival solution to calibrate the Amici spectrum in
wavelength. As was described in Fernandez-Soto et al (2009) we
needed to add an offset of 5 pixels, determined via comparison with
the observed sky absorption features. 

As we mentioned in the previous Section, there is a nearby star that
falls within the slit---it was in fact used to position the slit, as
the afterglow was too dim to be pointed at directly. Its spectral type
is M4V, as determined via available SDSS and 2MASS photometry
(Adelman-McCarthy et al 2008, Skrutskie et al 2006, see Table 1 for
the complete data). We have used the corresponding spectrum from the
Bruzual-Persson-Gunn-Stryker library (Bruzual et al 1996) to determine
the total efficiency of the instrument as a function of wavelength. 
In the $0.95-1.1\mu m$ region, corresponding to the Lyman break of a $z\approx 8$ source --- the most prominent feature in our GRB spectrum --- 
the spectrum of an M4V star is free from strong spectral features. This, considered
together with the low resolution of the spectra and the fact that at higher wavelengths the M4V star shows even less features, allows to rely on the total instrumental efficiency as derived by us.

It must be noted that there is a free factor involved in the
calculations, as we cannot ensure that the slit losses in the stellar
spectrum are the same in the one corresponding to the
afterglow. However, as long as the pointing was reasonably
accurate---and we can assume it was from the comparison of the
fluxes---at least to first order it will be a single number (i.e., not
wavelength-dependent) as the slit angle is obviously the same for both
objects.

The seeing at the time of the observations was $\approx 1.4$
arcseconds in the J band, as measured in the acquisition images. The
slit width was 1.0 arcsecond, hence we expect some slit losses, and no
degradation in resolution induced by the slit width. We show in Figure 3 
the measured total efficiency, obtained as the (arbitrarily normalised) ratio 
of the counts to the model spectrum of the star.

\begin{figure}
\centering
\includegraphics[angle=-0,width=8cm]{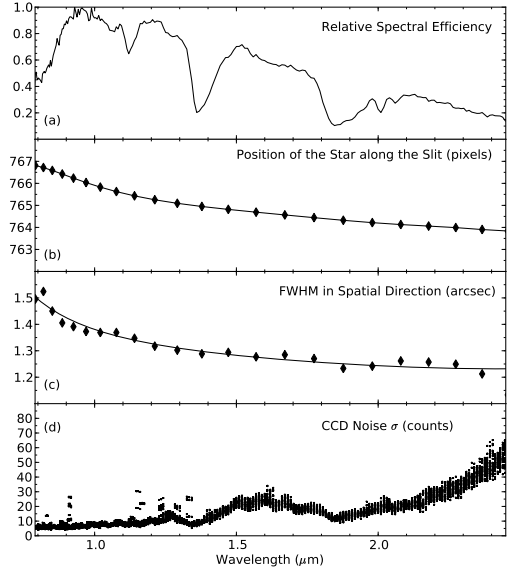}
\caption{Total efficiency of the atmosphere, telescope, and instrument
  combination (a). Changes in the spectral position (b) and width (c) with
  wavelength, and projection of the measured CCD noise (rms counts
  per pixel) as a function of wavelength (d).  }
      \label{FigStar}
\end{figure}

As is usually done when reducing spectroscopic data of very faint
sources, we will assume that the shape of the afterglow spectrum in
the CCD follows the trace left by the much brighter star (plotted in
Figure 3, second panel from the top), which can easily be traced at
all wavelengths from 0.8 to 2.5 microns. We determined the distortions
along the dispersion axis of the stellar spectrum, fitted it via
Legendre polynomials and used it to define the template spectrum
position. We determined the vertical offset by using 
a zero-order solution (an $f_{\nu}$-flat spectrum at $z=8.4$ with HI absorption of $10^{20}$cm$^{-2}$) and displacing it vertically, evaluating its likelihood at
different positions when compared to the CCD data. We measure an
offset of 115.4 pixels (equivalent to 28.8 arcseconds).

The FWHM in the spatial direction of the stellar spectrum varies with
wavelength, from $\approx 1.5$ arcseconds at $2.2 \mu$m to $\approx 1.2$ arcseconds at $\approx 0.8 \mu$m (corresponding to $\approx 5.8$ to $\approx 4.9$ pixels on our plate scale). We have used a smoothed fit to those
values (also shown in Figure 3) to reproduce the afterglow spectrum.

\subsubsection{CCD Characteristics}

The CCD image we are working with is the result of a careful reduction
procedure, done following the usual steps for NIR spectroscopy. We
have however performed one extra check: we measured the background
in detail, to ensure that it is flat in both the spectral and spatial
directions, and found no evidence of any trend in any of them, with
the background being flat to a fraction of the CCD noise. 

We also estimated the CCD noise using different methods. This is a
very important step, as we want to determine not only the basic
parameters of the GRB afterglow (redshift, spectral slope, and neutral
Hydrogen column density), but also to measure confidence limits on
both. We decided that the best method consists in using 7 rectangular
areas of equal size in the CCD, each one covering the interval $X \in
[480,650]$ and 21 pixels in the Y direction. Those boxes were chosen
in areas which are free of any (positive or negative) feature. Using
them we estimated the noise as a function of X position (that is,
wavelength) by using a 3x3 grid around each pixel and all 7
independent images. In this way we obtained a $170 \times 21$ mini-CCD
noise frame, that we will use for the subsequent chi-square
analysis. The projection of this noise array on the Y direction is
also shown at the bottom in Figure 3.

We finally computed the noise expected from an ideal $\chi^2\equiv1$ relation when comparing each one of the noise regions to an empty (zero counts) one. The average of the obtained noise values differs from the RMS of our noise array by $\approx 0.3\%$.

\begin{figure*}
\centering
\includegraphics[angle=-0,width=16cm]{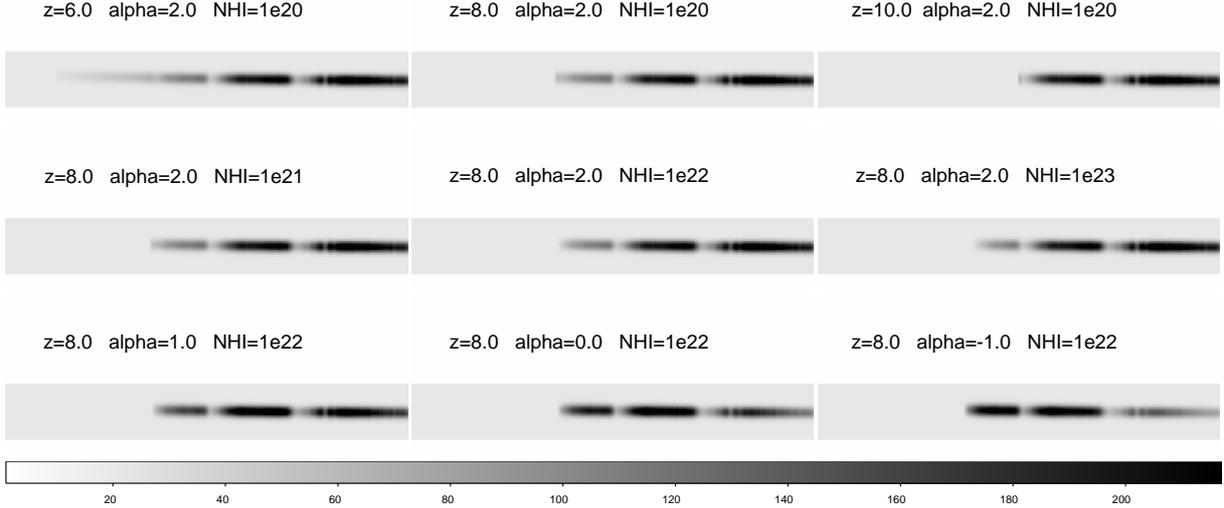}
   \caption{Some of the templates that were shown in Figure 1 are
     shown here as two-dimensional arrays, after application of the
     procedures described in Section 3.3.
           }
      \label{Fig2dspec}
\end{figure*}

\subsection{Application of the method}

We use all the knowledge about the instrument gathered in the previous
Subsection to create replicas of the GRB afterglow spectrum, for each
one of the templates described in Section 3.1. 

In brief, we choose a spectrum template combination (i.e., values of
$z$, $\alpha$, and N(HI)) and generate a one-dimensional spectrum
using them. We redden this spectrum using the measured Galactic value
E($B-V$)=0.03. Then this spectrum is converted to a two-dimensional
image, using the measured efficiency, adjusting and block-averaging
the wavelength axis to the known dispersion solution of the Amici
prism, and convolving it in the spectral direction with a Gaussian
function of the measured FWHM at each wavelength. This two-dimensional
spectrum is forced to follow the trace that was measured with the
stellar spectrum, and is set on a mini-CCD matrix measuring $170
\times 21$ pixels, as was described above for the noise image. Figure
4 shows some of the same spectra that were presented in Figure 2, now
converted to two-dimensional images with this process.

Each of them is then compared to the real data, using the section of
the CCD that contains the GRB afterglow spectrum, centered to the same
position of each of the template frames. We do this comparison using a
$\chi^2$ fit, where the noise array corresponds to the one that was
obtained in the previous section. Calling the template
$\mathcal{T}(z,\alpha,\rm{N(HI)})$, the CCD data $\mathcal{D}$, and
the noise matrix $\mathcal{S}$, and remembering each one of them
represents a $170 \times 21$ matrix, one obtains:
\begin{equation}
\chi^2[z,\alpha,\rm{N(HI)}] = \sum_{i=1}^{170} \sum_{j=1}^{21} 
  \frac{ [ A \mathcal{T}(z,\alpha,\rm{N(HI)})_{ij} - \mathcal{D}_{ij} ]^2}
       { \mathcal{S}_{ij}^2 },
\end{equation}
where $A$ represents a normalization parameter for the flux, which is
actually fixed for each template by minimising $\chi^2$, which renders
\begin{equation}
A=\frac{\sum_{i=1}^{170} \sum_{j=1}^{21} 
\mathcal{D}_{ij}
\mathcal{T}(z,\alpha,\rm{N(HI)})_{ij}/\mathcal{S}_{ij}^2}
{\sum_{i=1}^{170} \sum_{j=1}^{21} 
\mathcal{T}(z,\alpha,\rm{N(HI)})_{ij}^2/\mathcal{S}_{ij}^2}.
\end{equation}
Thus, once the normalisation $A$ is fixed, the calculation of $\chi^2$
is straightforward for each template.

\section{Results}

\begin{figure}
\centering
\includegraphics[angle=-0,width=9.5cm]{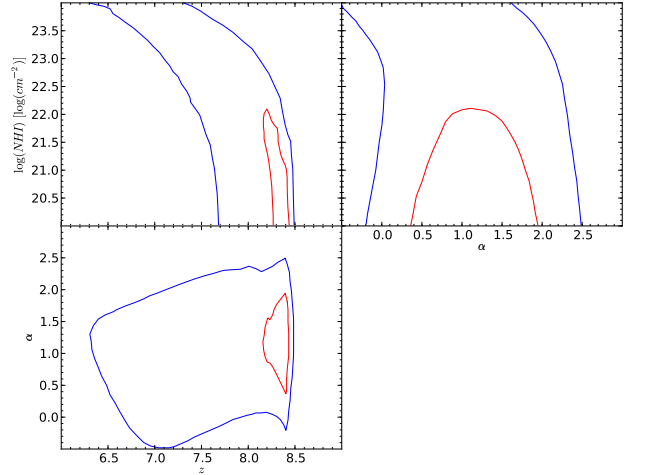}
   \caption{Results of the fitting procedure, projected on each of the 
   two-dimensional planes of parameter space. Red (blue) contours correspond to $1\sigma$ ($2\sigma$) confidence levels.
           }
      \label{FigChi2}
\end{figure}

In Figure 5 we present the result of the calculation above, projected
on the different planes of parameter space. Throughout our analysis, we associated the $68.3\%$ and $95.4\%$ confidence levels to $1\sigma$ and $2\sigma$ respectively. As can be seen, the best
fit corresponds to $z=8.40$ , $\alpha=1.2$,
$\rm{N(HI)}=10^{20.0}\rm{cm}^{-2}$. We have calculated the confidence
regions corresponding to the best fit on each of the individual parameters, reaching the
results presented in Table 2. We must remark that the best-fit
solution represents a value of $\chi^2 = 4019$, which indicates a
good fit for our problem, which has $\approx$ 3600 degrees of
freedom---albeit obviously most of them void in terms of information
content.

As can be observed there is no lower limit to the neutral hydrogen column density. 
This is a natural observational consequence of the fact that below 
N(HI)$\approx  2 \times 10^{20}$cm$^{-2}$ there is no damped profile, and the 
absorption is not significant at our resolution and signal-to-noise level. On the 
other end of the column density scale, we stop our analysis at 
N(HI)$= 10^{24}$cm$^{-2}$, a value large enough to include even the 
densest absorbers ever observed. Even larger values could be accomodated, 
paired to lower redshifts $z<6.5$ and flatter spectral slopes $\alpha <0.5$.

\begin{table}
\caption{Results of the fitting procedure. There is no valid 
limit to the HI column density at the $2\sigma$ level, which also 
somehow influences the other parameters (see text for further
details).}
\label{TabFits}
\centering
\begin{tabular}{l c c c}
\hline
    Parameter   & Value   &  $1\sigma$ & $2\sigma$ \\
\hline
Redshift          & 8.40 & (8.38,8,45)  & (6.67,8.49)  \\
Spectral Slope    & 1.2 & (0.7,1.8) & (0.1,2.2) \\
$\log$[N(HI)]     & 20.0 & ($<20.7$)  & --- \\
\hline
\end{tabular}
\end{table}

Figure 6 shows the best-fit spectrum, seen on the two top panels both as a one-dimensional
plot and as a clean two-dimensional model, and compared to the real data in the two lower
panels. 

\begin{figure}
\centering
\includegraphics[angle=-0,width=9.5cm]{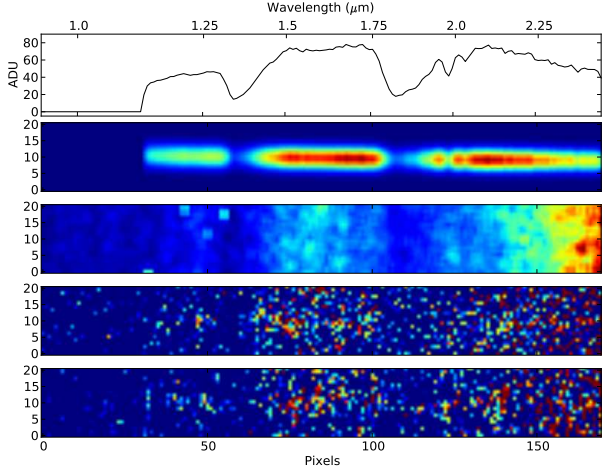}
   \caption{The best-fit spectrum, convolved with the instrument response, 
   is shown in the top panel as a one-dimensional spectrum, and as a clean 
   two-dimensional array in the second one. The third panel shows a $1\sigma$ 
   noise two-dimensional array in the CCD area corresponding to the afterglow 
   spectrum, as described in Section 3.2.2. The fourth panel shows one realization 
   of the CCD noise {\it plus} the best-fit spectrum, which can be directly compared 
   to the real data at the bottom panel. The contrast scale is the same in the two
   latter panels, but not in the previous ones.
           }
      \label{FigFin}
\end{figure}

We may also use the spectral slope of the GRB afterglow spectrum, as measured from broad-band NIR photometry, as a prior for our analysis. In this way, using the 1-sigma limit presented in Salvaterra et al (2009) for the spectral slope one would obtain a more stringent limit on the redshift, approximately $z>7.5$ at the 2-sigma confidence level.

\section{Conclusions}

We have presented a different method for the analysis of spectroscopic
data, akin in spirit to the one used in X-Ray astronomy and also in
photometric resdhift techniques. This method does not attempt to
extract a one-dimensional spectrum from the CCD data. On the other
hand, we generate simulated two-dimensional spectra from template
spectra of known characteristics, and decide which one fits best the
real data, choosing the best-fit parameters as the solutions to the
problem. 

We have used this method to perform a careful analysis of a very low
signal-to-noise spectrum of the afterglow of the very distant
GRB090423, showing that it is possible to extract information from
this kind of data, further than is usually assumed. Our best result
($z= 8.40^{+0.05}_{-0.02}$) is slightly higher than previous results obtained using the
same data (Salvaterra et al 2009) or data from other sources (Tanvir
et al 2009c), although still compatible well within the $2\sigma$ confidence level. 
We believe our confidence intervals to be reliable, given the full analysis 
of the noise characteristics performed in this work.

The same method can be applied to other cases where the source belongs
to a well-defined spectral class of objects, and there is a need to
extract the maximum possible information from low-quality data. This
is usually the case for faint, rapidly fading GRB afterglows, and we
expect this method to be used in future measurements.

\begin{acknowledgements}
  The authors gratefully acknowledge the support and effort of all members 
  in the CIBO collaboration ({\it Consorzio Italiano Burst Ottici}), particularly 
  Elisabetta Maiorano as PI of the proposal whence the Telescopio Nazionale
  Galileo GRB090423 data were obtained, and Stefano Covino for his help 
  with the manuscript. 

  AFS acknowledges support from the Spanish MICYNN projects
  AYA2006-14056 and Consolider-Ingenio 2007-32022, and from the
  Generalitat Valenciana project Prometeo 2008/132. It is also a
  pleaure to acknowledge the support of the Observatori Astronomic de
  la Universitat de Valencia during the development of this work.

This work is partly based on observations collected at TNG. The TNG
telescope is operated on the island of La Palma by the Centro Galileo
Galilei of the INAF in the Spanish Observatorio del Roque de Los
Muchachos of the Instituto de Astrofísica de Canarias.

This work made use of public data from the SDSS and 2MASS surveys.
The Two Micron All Sky Survey is a joint project of the University of
Massachusetts and the Infrared Processing and Analysis
Center/California Institute of Technology, funded by the National
Aeronautics and Space Administration and the National Science
Foundation. Funding for the SDSS and SDSS-II has been provided by the
Alfred P. Sloan Foundation, the Participating Institutions, the
National Science Foundation, the U.S. Department of Energy, the
National Aeronautics and Space Administration, the Japanese
Monbukagakusho, the Max Planck Society, and the Higher Education
Funding Council for England. The SDSS Web Site is {\tt
  http://www.sdss.org/}.  The SDSS is managed by the Astrophysical
Research Consortium for the Participating Institutions. The
Participating Institutions are the American Museum of Natural History,
Astrophysical Institute Potsdam, University of Basel, University of
Cambridge, Case Western Reserve University, University of Chicago,
Drexel University, Fermilab, the Institute for Advanced Study, the
Japan Participation Group, Johns Hopkins University, the Joint
Institute for Nuclear Astrophysics, the Kavli Institute for Particle
Astrophysics and Cosmology, the Korean Scientist Group, the Chinese
Academy of Sciences (LAMOST), Los Alamos National Laboratory, the
Max-Planck-Institute for Astronomy (MPIA), the Max-Planck-Institute
for Astrophysics (MPA), New Mexico State University, Ohio State
University, University of Pittsburgh, University of Portsmouth,
Princeton University, the United States Naval Observatory, and the
University of Washington.

\end{acknowledgements}

\end{document}